\documentclass[12pt,aps,prb,preprint]{revtex4}   

\usepackage{amsmath}    
\usepackage{amssymb}
\usepackage{graphicx}   

\usepackage[latin1]{inputenc}     
\usepackage[T1]{fontenc}

\usepackage{float}   

\draft

\begin{document}

\title{Proca equations derived from first principles}
\author{Michel Gondran}
 \affiliation{Université Paris Dauphine, 75 016 Paris, France.}
 \email{michel.gondran@polytechnique.com}   

\begin{abstract}
Gersten has shown how Maxwell equations can be derived from first
principles, similar to those which have been used to obtain the
Dirac relativistic electron equation. We show how Proca equations
can be also deduced from first principles, similar to those which
have been used to find Dirac and Maxwell equations.

Contrary to Maxwell equations, it is necessary to introduce a
potential in order to transform a second order differential
equation, as the Klein-Gordon equation, into a first order
differential equation, like Proca equations.

\end{abstract}

 \maketitle


\section{Introduction}

\bigskip

The Schrödinger and Klein-Gordon equations can be directly deduced
from first principles, i.e. from the condition liking the Energy
$E$, mass $m$, and momentum $\overrightarrow{\textbf{p}}$ and from
the correspondence principle where $E$ and
$\overrightarrow{\textbf{p}}$ are substituted by the quantum
operators $i \hbar \frac{\partial }{\partial t}$ and $ - i \hbar
\nabla$. The non relativistic condition $ (E-
\frac{\overrightarrow{\textbf{p}}^2}{2 m} - V) \psi=0$ where
$\psi$ is a wavefunction gives the Schrödinger equation and the
relativistic condition $(E^2 - c^2 \overrightarrow{\textbf{p}}^2 -
m^2 c^4) \psi=0 $ gives the Klein-Gordon equation.

The Dirac relativistic electron equation~\cite{Dirac_1928} is also
derived from first principles, but indirectly with a four
component wavefunction which permits to introduce a
spin-$\frac{1}{2}$ particle.

Gersten has shown~\cite{Gersten_1998} how Maxwell equations can be
obtained from first principles, similar to those which have been
used to infere the Dirac equation~\cite{Dirac_1928}.

The aim of the present paper is to show how Proca
equations~\cite{Proca_1936, Jackson_1962}, describing a massive
spin-1 particle, can be also derived from first principles, using
a decomposition similar to those which was used to find Dirac and
Maxwell equations. We can notice that the Proca equations can be
used for massive photons~\cite{Jackson_1962, Prokopec_2004} and
for the London penetration depth in a
superconductor~\cite{Jackson_1962, Ryder_2003}.

Contrary to Maxwell equations, it is necessary to introduce a
potential in order to transform a second order differential
equation, as the Klein-Gordon equation, into a first order
differential equation, as the Proca equations.

The method of Dirac and Gerster to obtained the Dirac and Maxwell
equations are recall in Section 2. The method to obtain the Proca
equations is described in Section 3.

\section{Dirac and Maxwell equations}

\bigskip

In his 1928 seminal paper~\cite{Dirac_1928}, Dirac deduces his
equation from the relativistic condition liking the Energy $E$,
mass $m$, and momentum $\overrightarrow{\textbf{p}}$:
\begin{equation}\label{eq:energieparticule}
(E^2 - c^2 \overrightarrow{\textbf{p}}^2 - m^2 c^4) I^{(4)}
\Psi=0,
\end{equation}
where $I^{(4)}$ is the 4 $\times$ 4 unit matrix and $\Psi$ is a
four component column (bispinor) wavefunction.

Dirac decomposes Eq. (\ref{eq:energieparticule}) into
\begin{equation}\label{eq:decompositionDirac}
[E I^{(4)} + \begin{pmatrix}
  m c^2 I^{(2)}  & c \overrightarrow{\textbf{p}} . \overrightarrow{\sigma} \\
  c \overrightarrow{\textbf{p}} . \overrightarrow{\sigma} & - m c^2 I^{(2)}
\end{pmatrix}][E I^{(4)} - \begin{pmatrix}
  m c^2 I^{(2)}  & c \overrightarrow{\textbf{p}} . \overrightarrow{\sigma} \\
  c \overrightarrow{\textbf{p}} . \overrightarrow{\sigma} & - m c^2 I^{(2)}
\end{pmatrix}]\Psi=0,
\end{equation}
where $I^{(2)}$ is the 2 $\times$ 2 unit matrix and
$\overrightarrow{\sigma}$ is the Pauli spin one-half vector matrix
with the components $\sigma_x=\begin{pmatrix}
  0 & 1 \\
  1 & 0
\end{pmatrix}$, $\sigma_y=\begin{pmatrix}
  0 & -i \\
  i & 0
\end{pmatrix}$ et $\sigma_z=\begin{pmatrix}
  1 & 0 \\
  0 & -1
\end{pmatrix}$.

\bigskip

Eq. (\ref{eq:decompositionDirac}) will be satisfied if the
equation
\begin{equation}\label{eq:decDirac}
[E I^{(4)} - \begin{pmatrix}
  m c^2 I^{(2)}  & c \overrightarrow{\textbf{p}} . \overrightarrow{\sigma} \\
  c \overrightarrow{\textbf{p}} . \overrightarrow{\sigma} & - m c^2 I^{(2)}
\end{pmatrix}]\Psi=0,
\end{equation}
will be satisfied. The Dirac equation will be obtained by
substitution in the equation (\ref{eq:decDirac}) of $E$ and
$\overrightarrow{\textbf{p}}$ by the quantum operators $i \hbar
\frac{\partial }{\partial t}$ and $ - i \hbar \nabla$.

\bigskip

The two component neutrino equation can be infered, in the same
manner, from the decomposition
\begin{equation}\label{eq:energiedecompositionneutrino}
[E^2 - c^2 \overrightarrow{\textbf{p}}^2) I^{(2)} \Psi=(E I^{(2)}-
c \overrightarrow{\textbf{p}}. \overrightarrow{\sigma}] [ E
I^{(2)} + c \overrightarrow{\textbf{p}}. \overrightarrow{\sigma}]
\Psi=0,
\end{equation}
where $\Psi$ is a two component spinor wavefunction.

\bigskip

Gersten~\cite{Gersten_1998} finds in 1998 the photon equation from
the relativistic condition on the Energy $E$ and momentum
$\overrightarrow{\textbf{p}}$:
\begin{equation}\label{eq:energiephoton1}
(\frac{E^2}{c^2} - \overrightarrow{\textbf{p}}^2) I^{(3)}
\overrightarrow{\Psi}=0,
\end{equation}
where $I^{(3)}$ is the 3 $\times$ 3 unit matrix and
$\overrightarrow{\Psi}$ is a three component column wavefunction.

Gersten decomposes Eq. (\ref{eq:energiephoton1}) into
\begin{equation}\label{eq:energiephoton}
 [\frac{E}{c}
I^{(3)} - \overrightarrow{\textbf{p }}.
\overrightarrow{\textbf{S}}][\frac{E}{c} I^{(3)} +
\overrightarrow{\textbf{p }}.
\overrightarrow{\textbf{S}}]\overrightarrow{\Psi} -
\begin{pmatrix}
 p_x\\
  p_y \\
  p_z
\end{pmatrix}
(\overrightarrow{\textbf{p}} . \overrightarrow{\Psi})= 0,
\end{equation}
where $\overrightarrow{\textbf{S}}$ is a spin one vector matrix
with components
\begin{equation}\label{eq:matricesspinun}
S_x=\begin{pmatrix}
  0  & 0  & 0 \\
  0  & 0  & -i \\
  0 & i & 0
\end{pmatrix},~~S_y=\begin{pmatrix}
  0  & 0  & i \\
  0  & 0  & 0 \\
  -i & 0 & 0
\end{pmatrix} ,~~S_z=\begin{pmatrix}
  0  & -i  & 0 \\
  i  & 0  & 0 \\
  0 & 0 & 0
\end{pmatrix} ,
\end{equation}
and with the properties
\begin{equation}\label{eq:eqcommutation}
[S_x, S_y]= i S_z, ~~[S_z, S_x]= i S_y, ~~[S_y, S_z]= i S_x,
~~\overrightarrow{\textbf{S}}^2= 2 I^{(3)}.
\end{equation}

\bigskip

Eq. (\ref{eq:energiephoton}) will be satisfied if the two
equations
\begin{equation}\label{eq:eqMaxwellphoton1}
[\frac{E}{c} I^{(3)} + \overrightarrow{\textbf{p }}.
\overrightarrow{\textbf{S}}]\overrightarrow{\Psi}=0,
\end{equation}
\begin{equation}\label{eq:eqMaxwellphoton2}
\overrightarrow{\textbf{p}} . \overrightarrow{\Psi}=0,
\end{equation}
will be simultaneously satisfied. The Maxwell equations will be
obtained by substitution in Eqs. (\ref{eq:eqMaxwellphoton1},
\ref{eq:eqMaxwellphoton2}) of $E$ and
$\overrightarrow{\textbf{p}}$ by the quantum operators $i \hbar
\frac{\partial }{\partial t}$ and $ - i \hbar \nabla$, and the
wavefunction substitution
\begin{equation}\label{eq:eqMaxwellphoton3}
\overrightarrow{\Psi} = \overrightarrow{\textbf{E}} - i
\overrightarrow{\textbf{B}},
\end{equation}
where $\overrightarrow{\textbf{E}}$ and
$\overrightarrow{\textbf{B}}$ are the electric and magnetic fields
respectively. With the identity
\begin{equation}\label{eq:identitegrad}
(\overrightarrow{\textbf{p }}.
\overrightarrow{\textbf{S}})\overrightarrow{\Psi}= \hbar \nabla
\times \overrightarrow{\Psi}
\end{equation}
Eqs. (\ref{eq:eqMaxwellphoton1}) and (\ref{eq:eqMaxwellphoton2})
give
\begin{equation}\label{eq:eqMaxwellimaginaire1}
i\frac{\hbar}{c}\frac{\partial(\overrightarrow{\textbf{E}} - i
\overrightarrow{\textbf{B}})}{\partial t}= - \hbar\nabla\times
(\overrightarrow{\textbf{E}} - i \overrightarrow{\textbf{B}}),
\end{equation}
\begin{equation}\label{eq:eqMaxwellimaginaire2}
\nabla . (\overrightarrow{\textbf{E}} - i
\overrightarrow{\textbf{B}})= 0.
\end{equation}
There are exactly all the Maxwell equations if the electric and
magnetic fields are real.
\bigskip

\section{Proca equations}

In this 1936 paper~\cite{Proca_1936}, Proca gives relativistic
wave equations for a massive spin-1 particle. We derive these
equations from the relativistic condition on the Energy $E$, mass
$m$, and momentum $\overrightarrow{\textbf{p}}$:
\begin{equation}\label{eq:energieboson}
(\frac{E^2}{c^2} - \overrightarrow{\textbf{p}}^2 - m^2 c^2)
I^{(3)} \overrightarrow{\Psi}=0,
\end{equation}
where $I^{(3)}$ is the 3 $\times$ 3 unit matrix and
$\overrightarrow{\Psi}$ is a three component column wavefunction.

Eq. (\ref{eq:energieboson}) can be decomposed into
\begin{equation}\label{eq:energiedecopossitionboson}
 [\frac{E}{c} I^{(3)} -
\overrightarrow{\textbf{p }}.
\overrightarrow{\textbf{S}}][\frac{E}{c} I^{(3)} +
\overrightarrow{\textbf{p }}.
\overrightarrow{\textbf{S}}]\overrightarrow{\Psi} -
\begin{pmatrix}
  p_x\\
  p_y \\
  p_z
\end{pmatrix}
(\overrightarrow{\textbf{p}} . \overrightarrow{\Psi}) - m^2 c^2
\overrightarrow{\Psi} =0,
\end{equation}
where $\overrightarrow{\textbf{S}}$ is a spin one vector matrix
defined by Eqs (\ref{eq:matricesspinun}) and
(\ref{eq:eqcommutation}).
\bigskip

Then we define the potential $(\varphi,
\overrightarrow{\textbf{A}})$ by the following equations:
\begin{equation}\label{eq:defpotentionboson1}
\overrightarrow{\textbf{p}} . \overrightarrow{\Psi}= i \frac{m^2
c^2}{\hbar}\varphi,
\end{equation}
\begin{equation}\label{eq:defpotentionboson2}
[\frac{E}{c} I^{(3)} + \overrightarrow{\textbf{p }}.
\overrightarrow{\textbf{S}}]\overrightarrow{\Psi}= i \frac{m^2
c^2}{\hbar}\overrightarrow{\textbf{A}}.
\end{equation}
If we use these two values in the
Eq.(\ref{eq:energiedecopossitionboson}) and substitute $E$ and
$\overrightarrow{\textbf{p}}$  by the quantum operators $i \hbar
\frac{\partial }{\partial t}$ and $ - i \hbar \nabla$, we find the
equation which decribe the wavefunction from the potential
\begin{equation}\label{eq:fonctondefonctionpotentiel}
\overrightarrow{\Psi}=- \frac{1}{c}\frac{\overrightarrow{\partial
\textbf{A}}}{\partial t}- i \nabla\times
\overrightarrow{\textbf{A}}- \nabla. \varphi.
\end{equation}

The other equations correspond to
Eqs.(\ref{eq:defpotentionboson1}) and
(\ref{eq:defpotentionboson2}) by substitution of $E$ and
$\overrightarrow{\textbf{p}}$ by the quantum operators $i \hbar
\frac{\partial }{\partial t}$ and $ - i \hbar \nabla$
\begin{equation}\label{eq:equationfoctionproca1}
\nabla \overrightarrow{\Psi}= - \frac{m^2 c^2}{\hbar^2}\varphi,
\end{equation}
\begin{equation}\label{eq:equationfoctionproca2}
\frac{1}{c}\frac{\partial \overrightarrow{\Psi}}{\partial t}- i
\nabla \times \overrightarrow{\Psi}=  \frac{m^2 c^2}{\hbar^2}
\overrightarrow{\textbf{A}}.
\end{equation}

If we put $\overrightarrow{\Psi} = \overrightarrow{\textbf{E}} - i
\overrightarrow{\textbf{B}}$, we deduce from
Eqs.(\ref{eq:fonctondefonctionpotentiel}),(\ref{eq:equationfoctionproca1})
and (\ref{eq:equationfoctionproca2}), if
$\overrightarrow{\textbf{E}}$ and $\overrightarrow{\textbf{B}}$
are real, the Proca equations
\begin{equation}\label{eq:EBfonctionpotentiel}
\overrightarrow{\textbf{E}}=- \frac{1}{c}\frac{\partial
\overrightarrow{\textbf{A}}}{\partial t}- \nabla.
\varphi,~~~~~~\overrightarrow{\textbf{B}}=\nabla\times
\overrightarrow{\textbf{A}}.
\end{equation}
\begin{equation}\label{eq:equationProca1}
\nabla . \overrightarrow{\textbf{E}}=-- \frac{m^2
c^2}{\hbar^2}\varphi,~~~~~~\nabla . \overrightarrow{\textbf{B}}=0,
\end{equation}
\begin{equation}\label{eq:equationProca2}
\frac{1}{c}\frac{\partial \overrightarrow{\textbf{E}}}{\partial t}
- \nabla \times \overrightarrow{\textbf{B}}=  \frac{m^2
c^2}{\hbar^2}
\overrightarrow{\textbf{A}},~~~~~~\frac{1}{c}\frac{\partial
\overrightarrow{\textbf{B}}}{\partial t}+ \nabla \times
\overrightarrow{\textbf{E}}= 0.
\end{equation}

\section{Conclusion}

Above, we have shown how all Proca equations, Eqs.
(\ref{eq:EBfonctionpotentiel}), (\ref{eq:equationProca1}) and
(\ref{eq:equationProca2}),  can be obtained from first principles,
similar to those which have been used to find Dirac and Maxwell
equations.

As demonstrated by Lévy-Leblond~\cite{Levyleblond_1967}, the spin
of the Dirac electron is not of relativistic origin, but primarily
a consequence of the use of a multicomponent wavefunction. It is
the same for the Maxwell and Proca particles.

The Dirac, Maxwell and Proca wavefunction are complex. Moreover
the Maxwell and Proca wave functions $\overrightarrow{\Psi} =
\overrightarrow{\textbf{E}} - i \overrightarrow{\textbf{B}}$ are
locally measurable and well understood quantities. Therefore these
wavefunctions should be used as a guideline for proper
interpretations of quantum theories.


\bigskip

\begin{thebibliography}{99}

\bibitem{Dirac_1928}
P. Dirac: "The Quantum Theory of the Electron", Proc. Roy. Soc.
\textbf{A117}, 620 (1928). (1928).


\bibitem{Gersten_1998}
A. Gersten: "Maxwell equations as the one-photon quantum
equation". Found. Phys. Lett. \textbf{12}, 291-8 (1998) .

\bibitem{Proca_1936}
A. Proca, Le Journal de Physique et le Radium \textbf{7}, 347
(1936).


\bibitem{Jackson_1962}
J. D. Jackson, \textit{Classical Electrodynamics}, John Wiley and
Sons, New York, 1999.

\bibitem{Prokopec_2004}
T. Prokopec and R. Woodard, "Vacuum polarization and photon mass
in inflation", Am. J. Phys. 72, 60-72 (2004).

\bibitem{Ryder_2003}
L. H. Ryder, \textit{Quantum Field Theory}, Cambridge University
Press, Cambridge, 2003.



\bibitem{Levyleblond_1967}
J. M. Lévy-Leblond, Comm. Math. Phys. \textbf{6}, 286 (1967).




\end{thebibliography}
\end{document}